\documentclass[aps,prd,twocolumn,showpacs,nofootinbib]{revtex4}
\usepackage{latexsym}
\usepackage{amsmath,amsfonts}
\usepackage{amsbsy}
\usepackage{mathrsfs}
\usepackage{color}

\usepackage{psfrag}

\usepackage{enumerate}

\usepackage{amsmath,amssymb,calc,amsfonts}
\usepackage{latexsym}

\usepackage{graphicx,calc,epsfig}
\def\ut#1{\rlap{\lower1ex\hbox{$\sim$}}#1{}}

\newcommand{\be}{\nopagebreak[3]\begin{equation}}
\newcommand{\ee}{\end{equation}}
\newcommand{\ba}{\nopagebreak[3]\begin{eqnarray}}
\newcommand{\ea}{\end{eqnarray}}
\DeclareFontFamily{U}{rsfs}{}         % Formal Script            %
\DeclareFontShape{U}{rsfs}{m}{n}{<5> rsfs5 <6><7> rsfs7          %
  <8><9><10><10.95><12><14.4><17.28><20.74><24.88> rsfs10}{}     %
\DeclareMathAlphabet{\mathfs}{U}{rsfs}{m}{n}                     %
\newcommand{\mfs}[1]{\mathfs {#1}}                               %
\newcommand{\n}{{\nonumber}}
\newcommand{\va}{\scriptscriptstyle}

\newcommand{\sI}{{\mfs I}}\newcommand{\sO}{{\mfs O}}

\newcommand{\nn}{\sqrt{j(j+1)}}

\def\pb#1{\rlap{\lower1.5ex\hbox{$\longleftarrow$}}{#1}}
\def\dpb#1{\rlap{\lower1.5ex\hbox{$\Longleftarrow$}}{#1}}
\def\spb#1{\rlap{\lower1.5ex\hbox{$\leftarrow$}}{#1}}
\def\sdpb#1{\rlap{\lower1.5ex\hbox{$\Leftarrow$}}{#1}}

\definecolor{blue}{rgb}{0,0,1}
\definecolor{green}{rgb}{0,1,0}
\definecolor{red}{rgb}{1,0,0}
\definecolor{vio}{rgb}{1,0,1}
\definecolor{ama}{rgb}{1,1,0}

\begin{document}

% \affiliation{$^2$Centre de Physique Th\'eorique\footnote{Unit\'e
% Mixte de Recherche (UMR 6207) du CNRS et des Universit\'es
% Aix-Marseille I, Aix-Marseille II, et du Sud Toulon-Var; laboratoire
% afili\'e \`a la FRUMAM (FR 2291)}, Campus de Luminy, 13288
% Marseille, France.}
%
%
% \affiliation{$^1$ \blue{FaMAF, Universidad Nacional de C\'ordoba,
%  Instituto de F\'isica Enrique Gaviola (IFEG), CONICET, \\
%  Ciudad Universitaria, (5000) C\'ordoba, Argentina.}
% }

%\maketitle

%\begin{abstract}
%\end{abstract}

%\pacs{}
\title{Black hole entropy and isolated horizons thermodynamics}

\date{\today}

\author{Amit Ghosh$^1$}
\author{Alejandro Perez$^2$}

\affiliation{$^1$Saha Institute of Nuclear Physics, 1/AF Bidhan Nagar,
700064 Kolkata, India.\\
$^2$Centre de Physique Th\'eorique\footnote{Unit\'e
Mixte de Recherche (UMR 6207) du CNRS et des Universit\'es
Aix-Marseille I, Aix-Marseille II, et du Sud Toulon-Var; laboratoire
afili\'e \`a la FRUMAM (FR 2291)}, Campus de Luminy, 13288
Marseille, France.}

%
%\affiliation{$^1$ . }

\begin{abstract}
We present a statistical mechanical calculation of the thermodynamical properties of (non rotating)
isolated horizons. The introduction of Planck scale allows for the definition of an
universal horizon temperature (independent of the mass of the black hole) and a well-defined
notion of energy (as measured by suitable local observers) proportional to the horizon area in
Planck units.  The microcanonical and canonical ensembles associated
with the system are introduced. Black hole entropy and other thermodynamical quantities can be consistently computed
in both ensembles and results  are in agreement with Hawking's semiclassical analysis for all values of the
Immirzi parameter.

\end{abstract}

%\pacs{}

\maketitle

%\section{Universal temperature at the Horizon}
Black holes are remarkably simple gravitational systems for distant observers so long as one neglects
quantum effects. However, for $\hbar\not=0$ their physical behaviour remains an open  question whose complete answer requires a full-fledged quantum gravity theory. The most difficult challenge is perhaps to unravel the physics close to the classical singularity dressed by the event horizon. The quantum gravity effects are also felt by observers outside the event horizon; as clearly indicated by Hawking's semiclassical calculations
\cite{Hawking:1974sw} which show that a generic BH radiates as a perfect black body at Hawking temperature $T_{\va H}$ proportional to its surface gravity and has an entropy $S=A/4\ell_p^2$ where $\ell_p=\hbar^{1/2}$ (in units $G=c=1$) is the Planck length and $A$ is the classical area of the event horizon. The analysis of these thermodynamic aspects of BHs is well within the reach of the existing developments in quantum gravity.

In fact, an account of the thermal properties of BHs from the statistical mechanical treatment of the  microscopic germs
arising in the underlying quantum theories of gravity has now become a standard benchmark for testing those theories. In this paper we attack this problem from the viewpoint of loop quantum gravity (LQG).

The problem of computing black hole entropy in the framework of LQG has a long history (see \cite{Corichi:2009wn} and references therein). Despite some small differences among various treatments, one common viewpoint has surely emerged which is that in order to find agreement with Hawking's semiclassical results one must fix the Immirzi parameter $\gamma$ (a dimensionless constant that labels various inequivalent kinematic quantizations of LQG) to a critical value $\gamma_0$. Although logically viable, this peculiar tuning of $\gamma$ has arguably become the Achilles' heel of the LQG analysis. In this paper we propose an alternative analysis of black hole entropy from LQG whose main merit is to reconcile Hawking's semiclassical results with the statistical mechanics treatment of LQG {\em without having to fix the Immirzi parameter}.

The key conceptual input is that the first law of black hole mechanics needs to be modified from the classical form $dE_{\infty}=\kappa dA/(8\pi)+\text{work terms}$ to
\be\label{firstlaw}
dE_{\infty}=\frac{\kappa}{8
\pi} dA+\mu_{\infty} dN+\text{work terms},
\ee
where $\kappa$ is the surface gravity of the event horizon, $E_{\infty}$ is the BH-mass measured by the stationary observers at infinity\footnote{In the usual statement of first Law $dE_\infty$ measures the mass difference of two different stationary black holes. Usually, one uses the notation $E_\infty=M$ where $M$ is the ADM-mass of the BH. Here, we use $E_{\infty}$ instead because we will also deal with a local version of the first law associated with the stationary observers in the interior of the spacetime for which the energy $E$ will differ from $E_{\infty}$.}
and the second term originates naturally from the underlying quantum geometry
description of the BH horizon where the integer $N$ refers to the number of topological
defects in the quantum isolated horizon (IH) and plays the role of a {\em quantum hair} for the BH. Then the quantity $\mu$ plays the role of chemical potential. As can be immediately seen, the above modification of the first law is fully consistent
with standard results for Schwarzschild BH if\footnote{In a static spacetime local temperature and chemical potentials are obtained from $T|g_{tt}|^{1/2}=T_\infty$ and
$\mu|g_{tt}|^{1/2}=\mu_\infty$, the local temperature is called the Unruh
temperature; see Landau and Lifshitz, Statistical Physics, Part I, $\S$27.}
$E_\infty=M$, $\mu_\infty=-\sigma(\gamma)T_\infty$ and
\be\label{pario}
S=\frac{A}{4\ell_p^2}+N \sigma(\gamma)
\ee
where $A$ is the classical area of IH and $\sigma(\gamma)$ is some function of the Immirzi parameter.
In the following we will show that the above first law and the entropy (\ref{pario}) follow directly
from the statistical mechanics of the basic quantum excitations of IH in LQG.
%We will also give an exact form of $\sigma(\gamma)$.
The Immirzi parameter is completely free and enters the entropy formula only through
the chemical potential.

In the following discussion, we use standard coordinates in which the Schwarzschild metric takes the form
\be
ds^2=-(1-\frac{2M}{r}) dt^2+{(1-\frac{2M}{r})^{-1}} {dr^2}+r^2 d\Omega^2.\label{metric}
\ee
One expects that in the semiclassical limit a spherically symmetric quantum black hole of large mass $M$ in a stationary state is well-approximated by a Schwarzschild BH having a test field in the Hartle-Hawking vacuum state. Equilibrium is sustained by a steady incoming flux of radiation at Hawking temperature $T_{\va H}={\ell_p^2}/({8\pi M})$ past null infinity $\sI^-$ and a steady outgoing flux of radiation
at the same temperature at future null infinity $\sI^+$. The temperature measured by a stationary observer in the interior is the local Unruh temperature
\be
T(r)={T_{\va H}}{({1-\frac{2M}{r}})^{-1/2}}\label{unruht}
\ee
Classically, this temperature diverges at the BH horizon. This is due to the infinite blue-shift of the asymptotic energy scales at the horizon. 
%Qualitatively one also (although one has to be more careful with this flat spacetime analogy) argue that the divergence is due to the infinite acceleration of the stationary observers at the horizon.
%What happens to free falling observers??? Nothing, in the Hartle-Hawking vacuum as well as in the Unruh
%vacuum the energy momentum tensor is regular at the future BH horizon. The BH horizon is still un-detectable by local observations in the semiclassical context (see Birrell-Davis 2d example for simplest explanation, 4d $<T_{\mu\nu}>$ expressions in Parker-Toms seem wrong).
However, in the quantum theory there is a universal (independent of mass $M$) local temperature at the horizon. More precisely, for an observer at  $r=2M+\epsilon$ the proper distance from the horizon is $\ell=2(2M\epsilon)^{1/2}$ and from (\ref{unruht}) the local temperature is
\be\label{tu}
T_{\va U}=\frac{\ell^2_p}{2\pi \ell}.
\ee
%
%This can be seen as follows: Take a stationary observer at $r=2M+\epsilon$. Then from (\ref{unruht})
%\be
%T(2M+\epsilon)=
%%\frac{T_{H}} {\sqrt{1-\frac{2M}{2M+\epsilon}}}=
%T\sqrt{\frac{2M}{\epsilon}}=\frac{\ell^2_{p}}{4 \pi}\sqrt{\frac{1}{2M\epsilon}}
%\ee
%where $\ell=(2M\epsilon)^{1/2}$ is the proper distance from the BH horizon.
Classically, $\ell\to 0$ as $\epsilon\to 0$. 
Quantum mechanically, the closest proper distance $\ell$ must be given by the smallest length scale that the quantum geometry can probe and hence it must be set by the underlying quantum theory of gravity. For example, in string theory $\ell$ must be determined by the string tension $\alpha'$; in LQG $\ell\sim\gamma\ell_p$. Remarkably, none of the physical result depends on this scale; so we do not fix it anywhere. Nevertheless, the existence of such a scale makes the local temperature (\ref{tu}) measured by a stationary observer closest to the BH horizon universal.
This is the relevant temperature for the quantum theory of isolated horizon and in its own spirit we  call it the {\em Unruh temperature}.

% \section{Quasilocal energy and horizon area}

The usual global notion of event horizon needs to be revised in the context of quantum gravity.
The very fact that BHs radiate in the semiclassical regime makes the definition of event horizon
as the boundary of the past of future null infinity unphysical; \cite{Ashtekar:2008jd} provides a clear description of this viewpoint in a simplified setting. In LQG this issue is resolved because one uses {\em isolated horizon}s \cite{Ashtekar:2004cn}. IH captures the main physical and local features of BH event horizons while being of a quasilocal nature itself. In particular, isolated horizons satisfy a quasilocal version of the first law \cite{aa}
\be
dE_{\va IH}=\frac{\kappa_{\va IH}}{8\pi} dA+\text{work terms},
\ee
where $E_{\va IH}$ is a suitable quasilocal energy function and $\kappa_{\va IH}$ is a local notion of IH surface gravity. Neither $E_{\va IH}$ nor $\kappa_{\va IH}$ are completely determined in the IH framework. More precisely, there are infinitely many possible first laws according to the choice of $\kappa_{\va IH}$ as a function of the extensive variables entering the first law which when integrated provides a definition of
$E_{\va IH}$.

This indeterminacy can only be eliminated by an appropriate physical input which makes an IH the closest representative of a BH spacetime. In the spherically symmetric case, this input is provided by the Schwarzschild spacetime. Indeed, there is a natural quasilocal energy that one can associate with the stationary observers in Schwarzschild spacetime. The four velocity of such an observer is $u^a=\xi^a/|\xi \cdot\xi|^{1/2}$ where $\xi$ the timelike killing vector field ($\xi=\partial_t$ in the coordinate system (\ref{metric})). Then the Komar mass integral gives
\be
E_r=-\frac{1}{8\pi} \int\limits_{S_r}  \epsilon_{abcd} \nabla^{c}u^d,
\ee
where $S_r$ is a spherical section of the $r=$ constant surface.
% located at a proper distance $\ell$ away from the event horizon.
The integral gives
 %%%%%%%%%%%%%%%%%%
 %\footnote{The explicit calculation is
 %\ba
 %E&=&-\frac{1}{8\pi} \int\limits_{S_{\epsilon}}  \epsilon_{abcd} \nabla^{c}u^d\n \\
  %&=&-\frac{1}{8\pi} \int\limits_{S_{\epsilon}}  \frac{\epsilon_{abcd} \nabla^{c}\xi^d}{\sqrt{-\xi\cdot\xi}}+\frac{1}{16\pi} \int\limits_{S_{\epsilon}}  \frac{\epsilon_{abcd} \xi^d \nabla^{c}(-\xi\cdot\xi)}{(\sqrt{-\xi\cdot\xi})^3} \n \\
  %&=& \frac{M}{\sqrt{1-\frac{2M}{r_{\epsilon}}}}+\frac{1}{16\pi} \int\limits_{S_{\epsilon}}  \frac{\epsilon_{abcd} \xi^d g^{cr} \partial_r(-\xi\cdot\xi)}{(\sqrt{-\xi\cdot\xi})^3} \n \\
  %&=& \frac{M}{\sqrt{1-\frac{2M}{r_{\epsilon}}}}-\frac{1}{2} \frac{{M}}{\sqrt{1-\frac{2M}{r_{\epsilon}}}} \n\\&=& \frac{M}{2\sqrt{1-\frac{2M}{r_{\epsilon}}}}= \frac{M^2}{\ell}
  %,
 %\ea
 %}
 %%%%%%%%%%%%%%%%%%
\be \label{eee}
E=\frac{M^2}{\ell}=\frac{A}{8\pi \ell},
\ee
when evaluated at $r=2M+\epsilon$ where $\ell=2\sqrt{2M\epsilon}$. This gives a natural notion of energy close to the BH horizon (see \cite{espera} for a full explanation of why this is the correct notion of energy for local observers). Since area is the only geometric quantity here, no wonder that the local energy is determined by the area where $16\pi\ell$ provides the appropriate scaling. The above analysis provides a clear-cut justification for the choice of area in the definition of microcanonical ensembles
used in \cite{Krasnov:1997yt}, and more profoundly recently in \cite{G.:2011zr}.

%\section{The quantum theory}

%We consider a gas of punctures in LQG. The universality of the temperature measured by observers flowing along the generators of the IH (see previous section) implies
%the relevant surface gravity that enters the the first law of IH mechanics is independent of the extensive variable describing the IH macroscopic geometry, i.e. its area $A$.
%From the first law
%\be
%dE(A)= \frac{\kappa}{8 \pi} dA
%\ee
%one concludes that the relevant notion of energy must be proportional to the IH area. As argued in the previous section we choose
%\be
%E(A)=\frac{A}{16\pi \ell}.
%\ee  Thus the dimensionless Immirzi parameter $\gamma$ is mediating the relationship between area and energy in our proposal.

%\subsection{The microcanonical ensemble}
From now on we study the statistical mechanical properties of quantum IHs.
As follows from the basic LQG treatment, we consider a quantum IH to be a gas of its topological defects, henceforth called {\em punctures}. Using (\ref{eee}), we take the appropriately scaled IH area spectrum as the energy spectrum of the gas. Using the LQG area spectrum \cite{lqg}
\be
\widehat H|j_1,j_2\cdots\rangle=(\frac{\ell^2_g}{2 \ell}  \sum_{p} \sqrt{j_p (j_p+1)})\  |j_1,j_2\cdots\rangle
\ee
where $j_p$ taking values from the set $\{1/2,1,3/2,...\}$ is the spin associated with the $p$-th puncture and we used the shorthand notation $\ell_g^2=\gamma\ell_p^2$.

The microcanonical ensemble is defined by an energy $E=A/(16\pi \ell)$ where $A$ is the classical area of the IH and a number of punctures $N$. A quantum configuration $\{s_j\}$ is given by the number of punctures $s_j$ carrying spin-$j$ for all possible values of $j$. Each configuration must obey two constraints
\begin{align}
&C_1:\;\sum_j\sqrt{j(j+1)}\,s_j=\frac{A}{8\pi\ell^2_g},\ \ \ \ 
&C_2:\;\sum_js_j=N.\n \end{align}
The number of states $d[\{s_j\}]$ associated with a configuration $\{s_j\}$ is
\be
d[\{s_j\}]={\Big(\sum_ks_k\Big)!}\prod_j\frac{(2j+1)^{s_j}}{s_j!}.
\ee
We look for the configuration that maximizes the entropy $\log(d[\{s_j\}])$ subject to the above two constraints. This configuration is obtained from the variational equation
\be
\delta\log(d[\{s_j\}])-\lambda\delta C_1-\sigma\delta C_2=0,
\ee
where $\lambda,\sigma$ are the two Lagrange multipliers. Under Stirling's approximation, this gives the dominant configuration
\be
\frac{s_j}{N}=(2j+1)e^{-\lambda\sqrt{j(j+1)}-\sigma}.
\ee
Summing over all spin values $j$ and using $C_2$, we get
\be
1=e^{-\sigma} \sum_j (2j+1) e^{-\lambda \sqrt{j(j+1)}}.
\ee
Denoting by $\bar d$ the value of $d[\{s_j\}]$ for the dominant configuration, the entropy $S=\log\bar d$ is given by
\begin{align}
&S=\lambda\frac{A}{8\pi\ell_g^2}+\sigma N
\quad\text{where}\nonumber\\
&\sigma(\gamma)=\log[\sum_j(2j+1)e^{-\lambda\sqrt{j(j+1)}}].
\end{align}
From $\beta=\left.{\partial S}/{\partial E}\right|_N$
we obtain Lagrange multiplier $\lambda$ as a function of $\beta$, namely $
\lambda=\beta{\ell_g^2}/({2\ell})$.
Finally, setting $T=T_{\va U}$ and using (\ref{tu}) we get
\begin{align}\label{micro}
&S= \frac{A}{4  \ell_p^2}+N\sigma(\gamma),\quad\text{where}\n\\
&\sigma(\gamma)= \log[\sum_j (2j+1) e^{-2 \pi\gamma  \nn }].
\end{align}
The function $\sigma(\gamma)$ appear at several places in what follows.
The chemical potential at $T=T_{\va U}$ is given by
\be\label{cp}
\mu=-T_{\va U}\left.\frac{\partial S}{\partial N}\right|_E=-\frac{\ell_p^2}{2\pi\ell}\,\sigma(\gamma)
\ee
which depends on the fiducial length scale $\ell$ and the Immirzi parameter.
% Let us introduce the notation
% \be
% \sigma(\gamma)=\log[\sum_j (2j+1) e^{-2\pi\gamma  \nn}]
% \ee

%\begin{figure}[htbp]
%\centering
%%\includegraphics[clip,width=0.8\textwidth]{blackhole1.pdf}
%\includegraphics[clip,width=0.4\textwidth]{chemical-potential.pdf}
%\caption{The chemical potential (in Planck units) as a function of $\gamma$. For small Immirzi parameter many punctures are
%energetically favoured. This  suggests a semiclassical limit dominated by complicated graphs and low spins.
%The opposite happens for large Immirzi parameter.
%}
%\label{fig:blhl1}
%`\end{figure}

%\subsection{The canonical ensemble}

For further discussion it is instructive to consider the same system in the canonical ensemble.
The canonical partition function is given by
\ba\label{zeta}
Z=\sum_{\{s_j\}}\prod_j\frac{N!}{s_j!}\,(2j+1)^{s_j}e^{-\beta s_jE_j}
%&=&\prod_{n} \exp((2j+1) e^{-\beta E_n})\n\\
%&=& \left(\sum_j (2j+1) e^{-\beta E_n}\right)^N,
\ea
where $E_j=\ell^2_g\nn/\ell$. A simple calculation gives
\be
\log Z=N\log[\sum_j(2j+1)e^{-\beta E_j}]
\ee
and the average energy 
$\label{energy}
\langle E\rangle=-\frac{\partial}{\partial\beta}\log Z
%\n \\
%&=& \frac{\frac{\ell_g^2}{2\ell} N e^{\frac{\ell_g^2}{4\ell} \beta}}{\sinh (\frac{\ell_g^2}{4\ell} \beta)+3 \cosh (\frac{\ell_g^2}{4\ell} \beta)-3}
$
at $T=T_{\va U}$ is a function of $N$ only; this relates the
number of punctures to the area
\be
N=-\frac{A}{4\ell_g^2\,\sigma'(\gamma)}.\label{NArelation}
\ee
Note that for all values of $\gamma$ the number of punctures $0\le N\le\frac{A}{4\sqrt 3\pi\ell_g^2}$. Moreover, for a fixed macroscopic area $A$, the number of punctures grows without limit as $\gamma\to 0$ while it goes to zero as $\gamma\to\infty$.
For the entropy we get
\begin{align}
\label{entro}
S=-\beta^2\frac{\partial}{\partial\beta}(\frac{1}{\beta}\log Z)=
% &=& \log(Z)+\beta \frac{\bar A}{16 \pi \ell} \n \\
\log Z+\beta\frac{A}{8 \pi\ell}.
%&=&N\left[\frac{1-3  e^{\frac{\ell^2_g}{4 \ell} \beta}+2 e^{\frac{\ell^2_g}{2\ell} \beta}+\frac{\ell^2_g}{4\ell}  \beta e^{\frac{\ell^2_g}{2\ell} \beta}}{(e^{\frac{\ell^2_g}{4 \ell} \beta}-1)^3}\right]
\end{align}
% We find
% \be
% S(N, A)=\frac{A}{4\ell^2_p}+N\sigma(\gamma).
% \ee
At $T=T_{\va U}$, this expression is identical to the micro-canonical entropy
(\ref{micro}). Basic formulae allow for the calculation of the energy fluctuations 
which 
%\be
%(\Delta E)^2=\langle E^2\rangle-\langle E\rangle^2=
%-\left.\frac{\partial \bar E}{\partial \beta}\right|_{N} =
%\left.\frac{\partial^2\log Z}{\partial\beta^2}\right|_N
%\ee
%and 
at the Unruh temperature are such that  
%\be
%%\frac{\ell_g ^4 N \left(3 e^{\frac{\beta \ell_g ^2}{4 \ell}}-2\right)}{8 \ell^2
%%   \left(\sinh \left(\frac{\beta \ell_g ^2}{4 \ell}\right)+3 \cosh \left(\frac{\beta
%%   \ell_g ^2}{4 \ell}\right)-3\right)^2}
%\ee
%from where we get
%\be
%\frac{\Delta^2 E}{\bar E^2}=\frac{e^{-\frac{\beta \ell_g ^2}{2 \ell}} \left(3 e^{\frac{\beta \ell_g ^2}{4
%   \ell}}-2\right)}{2 N}
%\ee
%which in thermal equilibrium at $T_{\va U}$ becomes
$
{(\Delta E)^2}/{\langle E\rangle^2}=\sO(1/N).
$
The specific heat at $T_{\va U}$ is
$
C={2N\ell\gamma}{\ell_p^{-2}}\sigma''(\gamma)
$
which is  positive. This implies that as a thermodynamic system the IH is locally
stable. The specific heat tends to zero in the large $\gamma$ limit for fixed
$N$ and diverges as $\hbar\to 0$.

%\section{Discussion}
We now conclude with some discussion of our results.

Hawking radiation is a global feature of spacetimes in which matter undergoes gravitational collapse and
settles down to some (semiclassical) stationary BH geometry. Quantum isolated horizons are ignorant about the geometry outside the horizon and hence are not expected to reproduce the thermodynamical properties of a global BH spacetime without additional inputs. Ideally, in LQG calculations, one should put by-hand the information of the semiclassical quantum states approximating the global BH geometry outside the IH as a physical input (such states are expected to exist in the large BH mass limit). Here we brought in some semiclassical inputs to the statistical treatment of quantum IH by setting the temperature of the IH at the appropriate blue-shifted Hawking temperature (\ref{tu}) and using the appropriate quasilocal energy (\ref{eee}). When these ingredients are incorporated into the entropy calculations, the consequences are striking.

In addition to the total area of the horizon, the total number of punctures
(topological defects on the membrane where surface degrees of freedom live)
is also a Dirac observable, which must play a role in the statistical description
of the horizon. 

Our result (20) for the entropy $S=\beta E+\sigma N$ is fully compatible
with semiclassical result of Bekenstein and Hawking even for a continuous
range of the Immirzi parameter. This follows form the fact that $(\partial
S/\partial E)_N=\beta$ must be the inverse temperature
(\ref{tu}) of the horizon and $(\partial S/\partial N)_E=\sigma$ must
be related to the chemical potential of the horizon. These suggest
that the correct first law of a quantum IH mechanics should be $dE=TdS+\mu dN$, where
the new term comes from the {\em quantum hair} $N$ that arises from the underlying
quantum geometry of IH, or more precisely from LQG. While comparing our
entropy with the semiclassical entropy one should note that that the latter is inferred
from an assumed form of the first law, i.e. the entropy $S$ has to be such that
$(\partial S/\partial E_\infty)_{...}=\beta_{\infty}$ holds, where the dots
refer to other possible macroscopic variables that must be kept fixed. Once one
equates $\beta_\infty=\beta_{\rm Hawking}$ and $E_\infty=E_{\rm ADM}$, one gets
the desired expression $S=A/4\ell_p^2$. Our entropy fully complies with this
analysis. The hair $N$, whose origin is purely quantum geometry, is held fixed
in the semiclassical analysis. Hence, the term $\sigma(\gamma)N$ is only an additive
constant to the entropy and at the semiclassical level, our entropy is the same
as the one of Bekenstein. This closes the gap between the semiclassical analysis
and the one of the statistical mechanics of IH.

Our chemical potential
$\mu=-T\sigma(\gamma)$ is negative for small values of $\gamma<\gamma_0$. So long as
$\mu\leqslant 0$, we can lower the energy of the horizon at some fixed entropy by
adding more punctures. That
means, large number of punctures is favoured. Also, for some fixed
energy, the entropy maximizes for maximum number of punctures. So large number
of punctures is also favoured entropycally. This shows that $N\gg 1$ is the
right semiclassical limit of geometry. Close to the value $\gamma_0$ of the
Immirzi parameter, the chemical potential tends to zero and for larger
values, it becomes positive. For $\gamma>\gamma_0$, a quantum theory may very well
exist mathematically, but it seems not to exhibit the right semiclassical behaviour.

The hair $N$ has its origin in the underlying quantum geometry and hence, the
first law of classical
isolated horizons do not possess this term. Classically, the only natural value
of the chemical potential is zero, which implies
$1=\sum(2j+1)\exp(-2\pi\gamma\nn)$. This fixes the value of the Immirzi
parameter reported earlier and from (20)  the entropy $S=A/4\ell_P^2$. This
result (with some mild differences depending up on the IH model) was obtained
in all previous counting \cite{countings}. Our present result can clearly
reproduce these earlier results. However, it differs in many important ways
from the existing viewpoint. First of all, in (20) the Immirzi parameter does
not appear as a multiplicative constant. It appears in an additive correction to
the semiclassical expression. This additive term is the quantum correction
to the semiclassical entropy induced by the quantum hair $N$. This result is
more robust in the sense that the semiclassical results are reproduced even when
$\gamma$ does not exactly obey the constraint and
the chemical potential is not exactly zero. Even to reproduce all earlier results
one only requires the chemical potential to be only close to zero; more precisely
$N\to\infty$ and $\sigma\sim\mfs O(1/N)$, so that the quantum correction to the
entropy $\sigma N\sim\mfs O(1)$.

%So the viewpoint is that the correct quantum statistical mechanics of isolated horizon should possess a non-zero chemical potential. This opens up a window for the Barbero-Immirzi parameter, $0<\gamma\leqslant \gamma_0$, for which we can still achieve the semi-classical result in the double limit, $N\to\infty$ and $\sigma\sim o(1/N)$, so that the quantum correction to the entropy $\sigma N\sim o(1)$. One can fine-tune the parameter to make the quantum correction as large as a logarithm of area and thus, cancel the leading log-correction to the entropy that arises from counting of states of the sub-dominant configurations. What we want to emphasize here that no such fine-tuning is needed to reproduce the correct semi-classical limit. This is one of the main points of this paper. (IS THE MU-GAMMA GRAPH RIGHT OR SHOULD IT BE INVERTED, I MEAN FOR SMALL VALUES OF GAMMA MU IS POSITIVE AND NEGATIVE FOR LARGE VALUES OF MU)

The quantum statistical mechanics of isolated horizon is independent of
the ensembles (we have shown the equivalence of microcanonical and canonical
ensembles). This is an important characteristic of a statistical
system in thermal equilibrium when some appropriate thermodynamic limit is
taken. In general, in absence of such a well-defined limit in gravity, one
expects that a black hole as a statistical system may exhibit features that
are ensemble dependent. Moreover, the thermodynamic description is ill-defined
because, for example for Schwarzschiled black hole, one finds the specific
heat is negative. For quantum isolated horizons, as we have shown here, nothing
is needed to overcome these difficulties.
%The equilibrium is naturally achieved.
The specific heat is positive and the system is in thermal equilibrium. This is
the main reason why we believe that this is the correct statistical description
of IH. The limit $N\to\infty$ plays the role of the thermodynamic limit in
our case (in other words, the semiclassical limit and the thermodynamic limit
are the same). 

Often the grand canonical ensembles provide more insights into the problem whose
partition function is $\mfs Z=\sum_0^\infty z^NZ(\beta,N)$ where $Z(\beta,N)$ is
the canonical partition function and $z=\exp(\beta\mu)$. It is not difficult to
see that $\mfs Z=[1-zf(T)]^{-1}$ where $f(T)=\sum(2j+1)\exp(-\beta E_j)$. The
average energy $\langle E\rangle$ and the average number of punctures $\langle N\rangle$ are related in the same way as (\ref{NArelation}). They also show that $zf(T)=1-\mfs O(1/N)$, so in the large $N$ limit and for $T=T_{\va U}$ the chemical potential is the same as (\ref{cp}). The entropy $S=\beta\langle E\rangle-\langle N\rangle\log z+\log\mfs Z$ is
\be S=S_{\rm micro}+\sO(\log N)
%+(N+1)\log(N+1)-N\log N
.\ee
Again, the deviation from the microcanonical entropy is small in the large $N$ limit. However, it is important to note that the fluctuations in $N$, and hence also in $E$, are $\mfs O(1)$. This signals to the fact that the system is in a phase transition region (see for example Pathria, Statistical Mechanics, 2nd Ed, $\S$4.5). This phase transition must have important significance in the quantum geometry description of IH. It suggests that a quantum IH exhibits critical behaviour.  This might imply that the semiclassical limit of IH is critical (similar to the continuum limit in lattice gauge theories where correlation lengths diverge). 
A different theoretical possibility is that this is  relevant for situations when the IH is  placed in the environment of other IHs with which it can exchange topological defects. Such environments arise during black hole mergers or quantum mechanical pair productions. The behaviour of IH differs significantly from the microcanonical or canonical descriptions in such situations. We keep this important issue of phase transition for future study.

Since the gas of punctures is in equilibrium at a high temperature (of
the order of Planck mass), statistically the punctures may very well be
bosonic or anyonic (the departure from Boltzmann statistics should be small). This may have
important implications at other temperatures, especially when the semiclassical approximation
breaks down and we have to deal with the quantum mechanics of punctures directly. We
wish to analyse this aspect of the gas also in the future.

The quantum corrections found here is different from the log-corrections that arise in all counting (the latter corrections arise from counting the sub dominant configurations and are also present in our model). A somewhat similar log-correction arise in the case of grand canonical ensemble and in that case the two corrections compete with each other. Investigations of this aspect is again kept for a future study.

% In conclusion, we propose that the quantum statistical mechanics of an IH is appropriately described by a gas of punctures in equilibrium at a local Unruh temperature. The only natural local energy scale set by gravity is (5), which is the energy measured by a static observer carrying the four-velocity (4) close to the horizon. With these minimal near-horizon input the quantum gas of punctures correctly reproduces the semi-classical entropy without having to fix the Immirzi parameter of the theory. The correction due to the quantum hair $N$ is additive and sub-leading. The fluctuations are studied in canonical ensembles. They show that in the semi-classical or in the thermodynamic limit the statistical behaviour of the system is independent of ensembles. The large values of the Immirzi parameter are not favoured because those quantum theories may not give the right semi-classical limit.

\begin{appendix}

\end{appendix}

\end{document}